\documentclass[useAMS]{mn2e}

\usepackage{latexsym,graphicx}

\usepackage{color}

\newcommand\kms{{\rm\,km\,s^{-1}}}
\newcommand\msun{\rm\,M_\odot}

\newcommand\hii{H\,{\sc ii} \,}

\def\apgt{\ {\raise-.5ex\hbox{$\buildrel>\over\sim$}}\ }
\def\aplt{\ {\raise-.5ex\hbox{$\buildrel<\over\sim$}}\ }

\title[$\zeta$ Oph and the weak-wind problem]{$\zeta$ Oph and the weak-wind problem}
\author[V.V.Gvaramadze et al.]
{V. V.~Gvaramadze,$^{1,2}$\thanks{E-mail: vgvaram@mx.iki.rssi.ru
(VVG); nlanger@astro.uni-bonn.de (NL); jmackey@astro.uni-bonn.de
(JM)} N.~Langer,$^{3\star}$ and J.~Mackey$^{3\star}$ \\
        $^{1}$Sternberg Astronomical Institute, Lomonosov Moscow State University, Universitetskij Pr. 13, Moscow 119992, Russia\\
        $^{2}$Isaac Newton Institute of Chile, Moscow Branch, Universitetskij Pr. 13, Moscow 119992, Russia\\
        $^{3}$Argelander-Institut f\"{u}r Astronomie, Universit\"{a}t Bonn, Auf dem H\"{u}gel 71, 53121 Bonn, Germany
        }

\begin{document}

\date{Accepted 2012 September 3. Received 2012 August 28; in original form 2012 July 25}

\maketitle

\label{firstpage}

\begin{abstract}
Mass-loss rate, $\dot{M}$, is one of the key parameters affecting
evolution and observational manifestations of massive stars, and
their impact on the ambient medium. Despite its importance, there
is a factor of $\sim 100$ discrepancy between empirical and
theoretical $\dot{M}$ of late-type O dwarfs, the so-called {\it
weak-wind problem}. In this Letter, we propose a simple novel
method to constrain $\dot{M}$ of runaway massive stars through
observation of their bow shocks and Str\"{o}mgren spheres, which
might be of decisive importance for resolving the weak-wind
problem. Using this method, we found that $\dot{M}$ of the
well-known runaway O9.5\,V star $\zeta$ Oph is more than an order
of magnitude higher than that derived from ultraviolet (UV)
line-fitting (Marcolino et al. 2009) and is by a factor of 6 to 7
lower than those based on the theoretical recipe by Vink et al.
(2000) and the H$\alpha$ line (Mokiem et al. 2005). The
discrepancy between $\dot{M}$ derived by our method and that based
on UV lines would be even more severe if the stellar wind is
clumpy. At the same time, our estimate of $\dot{M}$ agrees with
that predicted by the moving reversing layer theory by Lucy
(2010).
\end{abstract}


\section{Introduction}
%

%
\begin{figure*}
\includegraphics[width=17.5cm]{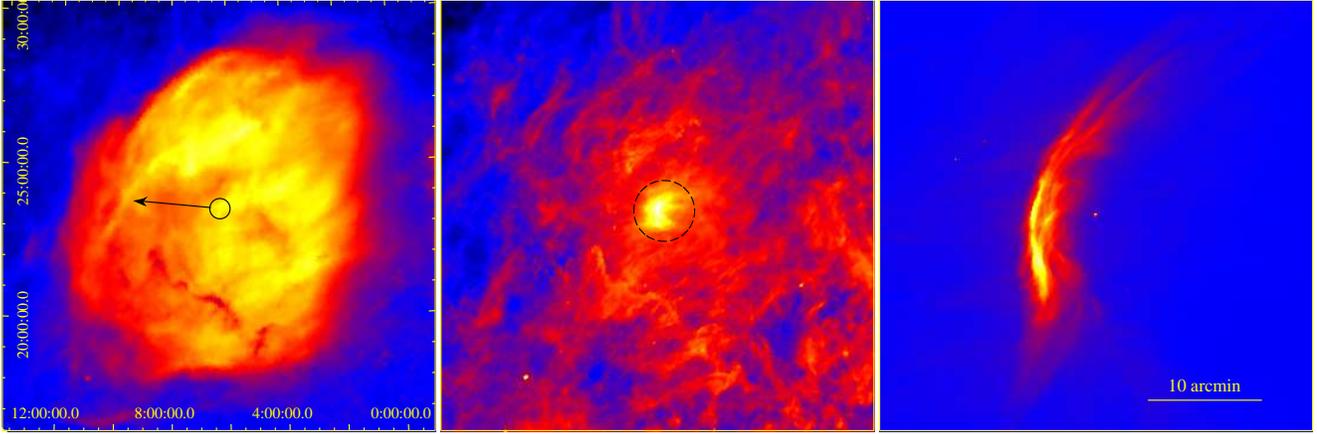}
\begin{center}
\caption{{\it Left}: SHASSA H$\alpha$ image of the H\,{\sc ii}
region Sh\,2-27. The position of the ionizing star, $\zeta$ Oph,
is marked by a circle, while the direction of its peculiar
(transverse) velocity is indicated by an arrow. The image is
oriented with Galactic longitude (in units of degrees) increasing
to the left and Galactic latitude increasing upward. {\it Middle}:
{\it IRAS} 60 $\mu$m image of the same field with the bow shock
generated by $\zeta$ Oph indicated by a dashed circle. The images
were generated by the NASA's SkyView facility (McGlynn, Scollick
\& White 1998). {\it Right}: {\it Spitzer} $24\,\mu$m image of the
bow shock around $\zeta$ Oph. The orientation of the images is the
same. At a distance of 112 pc, 1 degree corresponds to $\approx
1.93$ pc and 10 arcmin to $\approx 0.32$ pc.} \label{fig:hii}
\end{center}
\end{figure*}
\begin{table*}
\caption{Summary of astrometric and kinematic data on $\zeta$ Oph
(see text for details).} \label{tab:prop}
\begin{tabular}{cccccccc}
\hline
 $d$ & $\mu _\alpha \cos \delta$
& $\mu _\delta$ & $v_{\rm r,hel}$ & $v_{\rm l}$ & $v_{\rm b}$ & $v_{\rm r}$ & $v_*$ \\
 (pc) & (mas ${\rm yr}^{-1}$) & (mas ${\rm yr}^{-1}$) & ($\kms$) & ($\kms$) & ($\kms$) & ($\kms$) & ($\kms$) \\
\hline $112_{-2} ^{+3}$
& $15.26\pm0.26$ & $24.79\pm0.22$ & $-15.0$ & $26.4\pm0.1$ & $3.1\pm0.1$ & $-1.1$ & $26.5\pm0.1$ \\
\hline
\end{tabular}
\end{table*}

Mass-loss rate, $\dot{M}$, is one of the key parameters defining
the evolutionary sequence of massive stars and the end products of
their evolution (e.g. Chiosi \& Maeder 1986; Langer et al. 1994).
Although there is good agreement between the theoretically
predicted $\dot{M}$ of early-type O stars, i.e. stars more
luminous than $\approx 10^{5.2} L_{\odot}$, and those derived
empirically, there is a factor of 100 discrepancy between the
theoretical and empirical $\dot{M}$ of late-type O dwarfs (e.g.
Martins et al. 2005b, 2012; Marcolino et al. 2009), the so-called
{\it weak-wind problem} (see Hillier 2008 and Puls, Vink \&
Najarro 2008 for reviews).

$\dot{M}$ is also one of the main parameters defining the impact
of massive stars on the interstellar medium (ISM) through the
formation of stellar wind bubbles (e.g. Avedisova 1972; Castor,
McCray \& Weaver 1975) and bow shocks (Baranov, Krasnobaev \&
Kulikovskii 1971; Weaver et al. 1977). The characteristic scale of
these structures (typically tens of pc for bubbles and tenths of
pc for bow shocks) is proportional to $\dot{M}$ and therefore, in
principle, can be used to constrain this important parameter. The
use of the wind bubbles for this goal, however, is hampered
because most massive stars reside in parent clusters so that their
bubbles are the result of the joint action of all massive stars in
the cluster, which precludes us from constraining the wind
parameters of individual stars. On the other hand, wind bubbles
created by massive stars ejected into the field (the runaway
stars) rapidly turn into bow shocks (Weaver et al. 1977), whose
sizes depend on individual characteristics of their underlying
stars. This makes bow shocks an important tool for constraining
$\dot{M}$ of massive stars (Gull \& Sofia 1979; Kobulnicky,
Gilbert \& Kiminki 2010) and thereby for resolving the weak-wind
problem.

The characteristic size of a bow shock depends on the stellar wind
momentum rate (the product of $\dot{M}$ and the terminal velocity
of the stellar wind, $v_\infty$), the space velocity of the star,
and the number density of the ISM, $n$. The space velocities of
nearby stars can be accurately determined by measuring their
proper motions, parallaxes and radial velocities. Then $\dot{M}
v_\infty$ can be derived if $n$ is known.

In this Letter, we propose a simple novel method to constrain
$\dot{M}$ of massive stars based on detection of well-defined \hii
regions around bow-shock-producing runaways. Detection of these
structures around field O stars allows us to express $\dot{M}
v_\infty$ through observables (bow shock stand-off distance,
Str\"{o}mgren radius and space velocity) and the total
ionizing-photon luminosity, $S(0)$, which is a well-established
quantity for stars of known spectral type and luminosity class. We
applied this method to the well-known runaway O star $\zeta$ Oph,
which according to Marcolino et al. (2009) belongs to the group of
weak-wind stars (see Section\,\ref{sec:run} for a summary of
relevant data on this star). The excellent consensus in the
literature on $v_\infty$ of $\zeta$ Oph allowed us to estimate
$\dot{M}$ of this star separately, which turns out to be
intermediate between those based on $H\alpha$ and ultraviolet (UV)
lines (Section\,\ref{sec:constr}). The discussion of this result
and outlook are given in Section\,\ref{sec:dis}.

\section{$\zeta$ Oph, its bow shock and H\,{\sc ii} region}
\label{sec:run}

$\zeta$ Oph (HD\,149757, HIP\,81377) is a nearby ($\approx
112^{+3} _{-2}$ pc; van Leeuwen 2007), single (Garmany, Conti \&
Massey 1980), rapidly-rotating ($v\sin i\approx 400 \, \kms$;
Howarth \& Smith 2001), runaway (Blaauw 1961), O9.5\,Vnn (Morgan,
Code \& Whitford 1955; Lesh 1968) star with an impressive bow
shock (Gull \& Sofia 1979; van Buren \& McCray 1988) immersed in
the H\,{\sc ii} region Sh\,2-27 (Sharpless 1959).

In the left panel of Fig.\,\ref{fig:hii} we present the H$\alpha$
image of Sh\,2-27 originating from the Southern Hemispheric
H$\alpha$ Sky Survey Atlas (SHASSA; Gaustad et al. 2001), which
shows an almost circular ($\approx 10\degr$ or 9.6 pc in diameter)
\hii region. The middle panel of Fig.\,\ref{fig:hii} shows the
{\it Infrared Astronomical Satellite} ({\it IRAS}) 60 $\mu$m image
of the same field with an arcuate structure in the centre of the
\hii region, which is the bow shock generated by $\zeta$ Oph (van
Buren \& McCray 1988). Using the archival {\it Spitzer Space
Telescope} 24\,$\mu$m image (Program Id.: 30088, PI:
A.~Noriega-Crespo) of the bow shock (see the right panel of
Fig.\,\ref{fig:hii}), we estimated the angular separation between
the apex of the bow shock and the star of $\approx 5$ arcmin,
which corresponds to $\approx 0.16$ pc.

In Table\,\ref{tab:prop} we give the parallactic distance, $d$,
and the proper motion components, $\mu_\alpha \cos \delta$ and
$\mu_\delta$, of $\zeta$ Oph (all from the new reduction of the
{\it Hipparcos} data by van Leeuwen 2007), and the heliocentric
radial velocity, $v_{\rm r, hel}$, of this star (Wielen et al.
1999). To these data we added the components of the peculiar
transverse velocity (in Galactic coordinates), $v_{\rm l}$ and
$v_{\rm b}$, the peculiar radial velocity, $v_r$, and the total
space velocity, $v_* \equiv (v_l ^2 +v_b ^2 +v_{\rm r} ^2)^{1/2}$,
of the star. To calculate $v_*$, we used the Galactic constants
$R_0 = 8.0$ kpc and $\Theta _0 =240 \, \kms$ (Reid et al. 2009)
and the solar peculiar motion
$(U_{\odot},V_{\odot},W_{\odot})=(11.1,12.2,7.3) \, \kms$
(Sch\"onrich, Binney \& Dehnen 2010). For the error calculation,
only the errors of the proper motion measurement were considered.
It follows from Table\,\ref{tab:prop} that $\zeta$ Oph is moving
almost in the plane of sky.

The fundamental parameters of $\zeta$ Oph were studied in numerous
publications, of which the most recent are those by Mokiem et al.
(2005) and Marcolino et al. (2009). There is excellent consensus
on $v_\infty$ of the star: $1470 \, \kms$ (Prinja, Barlow \&
Howarth 1990), $1550 \, \kms$ (Repolust, Puls \& Herrero 2004) and
$1500 \, \kms$ (Marcolino et al. 2009). In what follows, we adopt
$v_\infty =1500 \, \kms$. Estimates of $\dot{M}$, however, are
very different. The theoretical recipe by Vink, de Koter \& Lamers
(2000) predicts $\dot{M} _{\rm Vink} \approx 1.29\times 10^{-7}
\msun \, {\rm yr}^{-1}$ (for stellar parameters derived by
Marcolino et al. 2009), which is comparable to $\dot{M} _{{\rm
H}\alpha} \approx 1.43\times 10^{-7} \msun \, {\rm yr}^{-1}$,
inferred by Mokiem et al. (2005) from synthesizing the H$\alpha$
line using the FASTWIND code. A value two orders of magnitude
lower, $\dot{M} _{\rm UV} \approx 1.58\times 10^{-9} \msun \, {\rm
yr}^{-1}$, was inferred by Marcolino et al. (2009) from CMFGEN
model fits to the UV doublet of C\,{\sc iv} $\lambda\lambda$1548,
1551, while Lucy (2010) predicted $\dot{M}_{\rm Lucy} \approx
1.30\times 10^{-8} \msun \, {\rm yr}^{-1}$ by using the updated
moving reversing layer theory of Lucy \& Solomon (1970).

The fast rotation of $\zeta$ Oph makes its spectral classification
somewhat ambiguous. Although the widely accepted spectral type is
O9.5\,V, some other classifications were suggested as well. For
instance, Conti \& Leep (1974) classified this star as O9\,V(e),
while Herrero et al. (1992) prefer O9\,III. Like Marcolino et al.
(2009), we adopt the spectral type of O9.5\,V, so that
$S(0)=3.63\times 10^{47} \, {\rm s}^{-1}$ (Martins, Schaerer \&
Hillier 2005b). The basic parameters of $\zeta$ Oph are summarized
in Table\,\ref{tab:sum}. To this table we added our estimate of
the mass-loss rate, $\dot{M} _{\rm obs}$, based on the observed
parameters of the bow shock and the \hii region associated with
$\zeta$ Oph (see next Section).

\section{Mass-loss rate of $\zeta$ Oph}
\label{sec:constr}

\begin{table*}
  \caption{Basic parameters of $\zeta$ Oph (see text for details).}
  \label{tab:sum}
  \begin{tabular}{ccccccccccc}
      \hline
      Spectral & $v_\infty$ & $S(0)$ & $\dot{M} _{\rm Vink}$ & $\dot{M} _{{\rm H}\alpha}$ & $\dot{M} _{\rm UV}$ & $\dot{M} _{\rm Lucy}$ & $\dot{M} _{\rm obs}$ \\
      type & ($\kms$) & (${\rm s}^{-1}$) & ($\msun \, {\rm yr}^{-1}$) & ($\msun \, {\rm yr}^{-1}$) & ($\msun \, {\rm yr}^{-1}$) & ($\msun \, {\rm
      yr}^{-1}$) & ($\msun \, {\rm yr}^{-1}$) \\
      \hline
       O9.5\,Vnn & 1500 & $3.63\times 10^{47}$ & $1.29\times 10^{-7}$ & $1.43\times 10^{-7}$ & $1.58\times 10^{-9}$ & $1.30\times 10^{-8}$ & $2.2\times 10^{-8}$ \\
       \hline
      \end{tabular}
   \end{table*}

The structure of an \hii region created by a moving star depends
on the stellar velocity relative to the local ISM. For
supersonically moving stars the number density within the \hii
region is comparable to that of the ambient ISM, while the radius
of the \hii region is of the order of the Str\"{o}mgren radius
(e.g. Tenorio Tagle, Yorke \& Bodenheimer 1979), which is given by
(e.g. Lequeux 2005)
\begin{equation}
R_{\rm St} = \left({3S(0) \over 4\pi \alpha _{\rm B}
n^2}\right)^{1/3} \, , \label{eqn:str}
\end{equation}
where $\alpha _{\rm B}$ is the recombination coefficient of
hydrogen to all but the ground state ($\approx 2.6\times 10^{-13}
\, {\rm cm}^3 {\rm s}^{-1}$ for fully ionized gas of temperature
of $10^4$ K).

The stellar wind of a moving star interacts with the ISM and
produces a bow shock ahead of the star. The minimum distance from
the star at which the wind pressure is balanced by the ram and the
thermal pressures of the ISM, the stand-off distance, is given by
(e.g. Baranov et al. 1971)
\begin{equation}
R_0 =\left[{\dot{M} v_\infty \over 4\pi n(\mu m_{\rm H}v_* ^2
+2kT)}\right]^{1/2} \, , \label{eqn:bow}
\end{equation}
where $\mu =1.4$ is the mean molecular weight, $m_{\rm H}$ is the
mass of a hydrogen atom, $k$ is the Boltzmann constant, and $T
(\approx 10^4$ K) is the temperature of the (ionized) ISM.

Eliminating $n$ between equations\,(\ref{eqn:str}) and
(\ref{eqn:bow}), one has the `observed' stellar wind momentum rate
\begin{eqnarray}
\dot{M} _{\rm obs} v_\infty =1.57\times 10^{25} \, {\rm g} \, {\rm
cm} \, {\rm s}^{-2} \left(1+{1\over M^2}\right) \left({R_0 \over
0.1 \, {\rm pc}}\right)^2 \nonumber \\
\times \left({v_* \over 10 \, \kms}\right)^2 \left({S(0) \over
10^{48} \, {\rm s}^{-1}}\right)^{1/2} \left({R_{\rm St} \over 10
\, {\rm pc}}\right)^{-3/2} \, , \label{eqn:mom}
\end{eqnarray}
where $M=v_* /c_{\rm s}$ is the isothermal Mach number and $c_{\rm
s}=(2kT/\mu m_{\rm H})^{1/2}$ is the isothermal sound speed
($\approx 10.9 \, \kms$ for $T=10^4$ K). For $R_0 =0.16$ pc, $v_*
=26.5 \, \kms$, $S(0)=3.63\times 10^{47} \, {\rm s}^{-1}$, and
$R_{\rm St} = 9.6$ pc, one has from equation\,(\ref{eqn:mom}) that
$\dot{M} _{\rm obs} v_\infty \approx 2.1\times 10^{26} \, {\rm g}
\, {\rm cm} \, {\rm s}^{-2}$. Thanks to the excellent consensus on
$v_\infty$ of $\zeta$ Oph (see Section\,\ref{sec:run}), one finds
$\dot{M}$ of this star separately, $\dot{M} _{\rm obs} =2.2\times
10^{-8} \, \msun \, {\rm yr}^{-1}$. This is $\approx 14$ times
larger than $\dot{M} _{\rm UV}$, 7 times lower than $\dot{M}
_{{\rm H}\alpha}$, and comparable to $\dot{M} _{\rm Lucy}$. The
discrepancy between $\dot{M} _{\rm UV}$ and $\dot{M} _{\rm obs}$
would be even more severe if the wind is clumpy. For instance,
$\dot{M} _{\rm UV}$ should be reduced by a factor of three ($\sim
\sqrt{f}$) if the volume filling factor, $f$, in the wind is $\sim
0.1$ (Marcolino et al. 2009), so that $\dot{M} _{\rm obs} \approx
40\dot{M} _{\rm UV}$.

Using equations\,(\ref{eqn:bow}) and (\ref{eqn:mom}), one finds
$n\approx 3.6 \, {\rm cm}^{-3}$, which agrees well with estimates
of the electron number density within Sh\,2-27 of $\approx 3.0 \,
{\rm cm}^{-3}$ (based on radio observations, H$\alpha$ surface
brightness, and studies of interstellar absorption lines; Gull \&
Sofia 1979 and references therein) and the average line of sight
electron and neutral hydrogen densities towards $\zeta$ Oph of
$\approx 4.0 \, {\rm cm}^{-3}$ (Howk \& Savage 1999; scaled to
$d$=112 pc).

Note that $\dot{M} _{\rm obs}$ was derived for $d=112$ pc, while
Marcolino et al. (2009) adopted a somewhat larger distance of 146
pc. An even larger value, $d=222$ pc, was suggested by Megier et
al. (2009), whose estimate is based on the empirical relationship
between the strength of the interstellar Ca\,{\sc ii} lines and
the distances to early-type stars. Since $v_*$ scales with the
distance as $d^{0.6}$ ($v_* \approx 31.2$ and $41.7 \, \kms$ for
$d=146$ and 222 pc, respectively), one finds that $\dot{M}_{\rm
obs} \propto d^{1.7}$, so that the larger the distance the larger
the discrepancy between $\dot{M}_{\rm obs}$ and $\dot{M}_{\rm
UV}$, namely $\dot{M} _{\rm obs} \approx 22-45 \dot{M} _{\rm UV}$
for $d$ in the range from 146 to 222 pc. Correspondingly, $\dot{M}
_{{\rm H}\alpha}$ remains larger than $\dot{M} _{\rm obs}$ by a
factor of $\approx 2$ to 4. $\dot{M} _{\rm Vink}$ also depends on
the adopted distance (via the stellar luminosity; Vink et al.
2000); this would decrease (increase) $\dot{M} _{\rm Vink}$ by a
factor of a few for $d=112$ (222) pc.

Similarly, it follows from equation\,(\ref{eqn:mom}) that
$\dot{M}_{\rm obs} \propto S(0)^{1/2}$, so that $\dot{M} _{\rm
obs} \approx 21 \dot{M} _{\rm UV}$ (or $\approx 0.2 \dot{M} _{{\rm
H}\alpha}$) if $\zeta$ Oph is an O9\,V star (Conti \& Leep 1974),
and $\dot{M} _{\rm obs} \approx 37 \dot{M} _{\rm UV}$ ($\approx
0.4 \dot{M} _{{\rm H}\alpha}$) if its spectral type is O9\,III
(Herrero et al. 1992). Here we used the calibrations of Martins et
al. (2005b).

On the other hand, the possible leakage of ionizing photons from
the \hii region (caused by porosity of the ISM; see Wood et al.
2005 and Section\,\ref{sec:dis}) would reduce our estimate of
$\dot{M} _{\rm obs}$. Conservatively assuming that a half of the
photons escape from Sh\,2-27, one finds that $\dot{M} _{\rm obs}$
should be reduced by a factor of $\approx 1.4$.

Note also that the possible existence of large-scale internal
flows in the \hii region (e.g. caused by photoevaporation of
molecular clumps in the ISM; see Section\,\ref{sec:dis}) also can
affect the estimate of $\dot{M} _{\rm obs}$ because $R_0$ depends
on the stellar velocity relative to the local ISM. Assuming that
the characteristic velocity of the photoevaporation flows is of
order of the sound speed in the ionized gas, i.e. $\sim 10 \,
\kms$, one finds that $\dot{M} _{\rm obs}$ might either be larger
or smaller by a factor of up to $\approx 2$ if the flow velocity
is antiparallel or parallel to the vector of the space velocity of
$\zeta$ Oph.

Finally, we note that the fast rotation of $\zeta$ Oph might make
the stellar wind anisotropic by increasing $\dot{M}$ in the
equatorial zone (Friend \& Abbott 1986). This in turn might affect
the geometry of the bow shock and the UV line and H$\alpha$
diagnostic. The H$\alpha$ diagnostic would also be unreliable if
the Oe status of $\zeta$ Oph is due to the presence of a
circumstellar disk (see, however, Vink et al. 2009). The
anisotropy effects cannot be easily estimated, but are unlikely to
change our result significantly.

To summarize, the uncertainties in the distance and the spectral
type might lead to the increase of $\dot{M} _{\rm obs}$ by a
factor of $\approx 1.5$ to 3, while the leakage of ionizing
photons from the \hii region and the presence of regular (e.g.
photoevaporation) flows within the \hii region might reduce
$\dot{M} _{\rm obs}$ by a factor of two. From this it follows that
$\dot{M} _{\rm obs}$ would remain at least an order of magnitude
larger than $\dot{M} _{\rm UV}$ and might be comparable to or
somewhat lower than $\dot{M} _{{\rm H}\alpha}$.

\section{Discussion and further work}
\label{sec:dis}

We estimated the mass-loss rate, $\dot{M}_{\rm obs}$, of the
candidate weak-wind star $\zeta$ Oph using the observed parameters
of its bow shock and \hii region. We found that $\dot{M}_{\rm
obs}$ is more than ten times larger than $\dot{M}_{\rm UV}$
inferred by Marcolino et al. (2009). (Recall that the difference
between the two estimates would be even higher if the stellar wind
is clumpy.) This finding supports the suggestion by Mokiem et al.
(2007) that the use of only UV lines might significantly
underestimate $\dot{M}$ of late-type O stars, because X-rays
created by shocked wind regions can significantly change the wind
ionization and thereby reduce the wind emission in the UV lines
(see Martins et al. 2005a and Marcolino et al. 2009 for more
details). Moreover, the strengths of diagnostic UV lines could
also be reduced (for a given $\dot{M}$) if the wind is porous
(clumpy) not only spatially, but also in velocity space
(Sundqvist, Puls \& Feldmeier 2010; Muijres et al. 2011). On the
other hand, although the latter effect does not significantly
affect the H$\alpha$ line, the H$\alpha$ based estimates of
$\dot{M}$ should be considered as upper limits because of the wind
clumpiness (Repolust et al. 2004).

The latter conclusion is in line with our finding that
$\dot{M}_{\rm obs}$ is $\approx 7$ times lower than $\dot{M}_{{\rm
H}\alpha}$ (derived by Mokiem et al. 2005 using the FASTWIND
code). Note that the discrepancy between the two estimates could
also be caused by the fact that the FASTWIND code tends to predict
a stronger H$\alpha$ absorption than the CMFGEN one (Puls et al.
2005; Marcolino et al. 2009), which can affect the H$\alpha$ based
estimates of $\dot{M}$. It is plausible therefore that the actual
$\dot{M}$ of $\zeta$ Oph is indeed somewhere between $\dot{M}_{\rm
UV}$ and $\dot{M}_{{\rm H}\alpha}$, as suggested by the work of
Lucy (2010) and as found in this Letter. This inference along with
the recent finding by Muijres et al. (2012) that the recipe by
Vink et al. (2000) might over-predict $\dot{M}$ for late-type O
dwarfs provides an avenue for resolving the weak-wind problem.

Our estimate of $\dot{M}_{\rm obs}$ was obtained under the
assumption that $\zeta$ Oph is running through a homogeneous ISM.
The [N\,{\sc ii}]/H$\alpha$ and [S\,{\sc ii}]/H$\alpha$ line ratio
maps of Sh\,2-27 (based on the Wisconsin H$\alpha$ Mapper
observations; Wood et al. 2005), suggest, however, that the ISM is
porous, i.e. contains low-density voids. One might wonder
therefore whether $\zeta$ Oph is currently located within such a
void so that its $\dot{M}$ is actually $\ll \dot{M} _{\rm obs}$
and correspondingly comparable to $\dot{M} _{\rm UV}$.

To address this issue we note that the ionizing radiation of
$\zeta$ Oph could homogenize the ambient ISM by photoevaporation
of density inhomogeneities (e.g. Elmergreen 1976; McKee, van Buren
\& Lazareff 1984). Thus, our assumption of a homogeneous medium
would be correct if the star completely photoevaporates clumps
within a region of radius $R_{\rm ph} \geq R_0$ on a time-scale,
$t_{\rm ph}$, shorter than the crossing time of this region, i.e.
$t_{\rm ph} < R_{\rm ph}/v_*$. The photoevaporation time of a
clump of mass $M_{\rm cl}$ located at a distance $R_{\rm ph}$ is
given by (McKee at al. 1984)
\begin{eqnarray}
t_{\rm ph} =1.7\times 10^4 \, {\rm yr} \left({n_{\rm m} \over 1 \,
{\rm cm}^{-3}}\right) \left({S(0)\over 10^{48}  \, {\rm
s}^{-1}}\right)^{-1/2} \nonumber \\ \times \left({M_{\rm cl} \over
\msun}\right)^{1/2} \left({R_{\rm ph} \over 1 \, {\rm pc}}\right)
\left({c_{\rm s} \over 10.9 \, \kms}\right)^{-1} \, ,
\end{eqnarray}
where $n_{\rm m}$ is the mean density the local ISM would have if
it were homogenized. The quite smooth appearance of the bow shock
(whose transverse dimension is $\approx 1$ pc) implies that the
local ISM is homogeneous on a scale of $\sim 1$ pc, so that our
assumption would be correct if the mass of individual clumps is
$\leq 0.01 \, \msun$ (we assumed here that $R_{\rm ph} =1$ pc and
$n_{\rm m} \approx n=3.6 \, {\rm cm}^{-3}$). It is difficult to
make more quantitative arguments analytically, because this is a
very non-linear process. We are currently using 3D
radiation-magnetohydrodynamics simulations to investigate the
effects of a clumpy ISM in more detail (Mackey et al., in
preparation).

The current sample of Galactic candidate weak-wind stars (Martins
et al. 2005, 2012; Marcolino et al. 2009) contains 22 stars
(including $\zeta$ Oph). We searched for bow shocks around all
these stars using the Mid-Infrared All Sky Survey carried out with
the {\it Wide-field Infrared Survey Explorer} ({\it WISE}; Wright
et al. 2010). This survey provides images in four wavebands
centred at 3.4, 4.6, 12 and 22\,$\mu$m, of which the 22\,$\mu$m
band is most suitable for detection of bow shocks (e.g. Gvaramadze
et al. 2011; Peri et al. 2012). Besides $\zeta$ Oph, the bow
shocks were detected around four stars: HD\,34078, HD\,48099,
HD\,48279, and HD\,216898. The first two stars are well-known
runaways (see figs.\,2 and 3 in Peri et al. 2012 for the {\it
WISE} images of their bow shocks), while the runaway status of
HD\,48279 (suggested by its space velocity of $\approx 30 \, \kms$
and the presence of a bow shock) was recently established by
Gvaramadze et al. (2012)\footnote{It is believed that HD\,48279 is
a member of the Mon\,OB2 association, while the orientation of the
bow shock and the vector of the stellar peculiar transverse
velocity imply that the star was {\it injected} into the
association (cf. Gvaramadze \& Bomans 2008).}. The bow shock
around HD\,216898 is detected for the first time.

Unfortunately, none of the four stars are associated with
well-defined \hii regions. There are, however, several alternative
possibilities to estimate the local ISM density. For example, the
pre-shock density could be derived through modelling (optical)
spectra of the shocks (e.g. Dopita 1977) or from their emission
measures, either in H$\alpha$ or in the infrared (e.g., Bally et
al. 2006; Kobulnicky et al. 2010). The latter possibility is
especially attractive because bow shocks are detectable mostly in
the infrared. Hydrodynamic modelling of bow shocks can also
constrain the ISM density by comparing observed instabilities, bow
shock mass, and emissivity to simulations (e.g. Mohamed, Mackey \&
Langer 2012). Further work in this direction is highly desirable
in order to constrain $\dot{M}$ of a larger sample of `weak-wind'
stars.

The paucity of bow-shock-producing `weak-wind' stars with
well-defined \hii regions can be compensated by detection of new
examples of late O/early B type stars possessing these structures.
Indeed, inspection of the SHASSA survey revealed several circular
\hii regions with central bow-shock-producing OB stars (Gvaramadze
et al., in preparation), some of which (e.g. $\delta$ Sco, $\tau$
Sco) might belong to the group of the `weak-wind' stars.
Determination of $\dot{M}$ for these stars using different
atmosphere models and their comparison with $\dot{M} _{\rm obs}$
(and $\dot{M}$ based on the work of Lucy 2010) is highly desirable
as well.

To conclude, detection of bow shocks (and well-defined \hii
regions) around field late-type O dwarfs provides unique
possibilities for resolving the weak-wind problem, which in turn
would have profound consequences for better understanding the
mass-loss mechanism of massive stars, their impact on the ambient
ISM, and their evolution in general.

\section{Acknowledgements}

We are grateful to the referee for useful comments. JM is funded
by a fellowship from the Alexander von Humboldt Foundation. This
work has made use of the NASA/IPAC Infrared Science Archive, which
is operated by the Jet Propulsion Laboratory, California Institute
of Technology, under contract with the National Aeronautics and
Space Administration, the SIMBAD database and the VizieR catalogue
access tool, both operated at CDS, Strasbourg, France.

\end{document}